\documentclass[prd,twocolumn,a4paper,superscriptaddress,floatfix]{revtex4}
\usepackage{graphicx}
\usepackage{bbm}
\usepackage[all]{xy}
\usepackage{amsmath}
\usepackage{amssymb}
\usepackage{epstopdf}

\def\be{\begin{equation}}
\def\ee{\end{equation}}
\newcommand{\bq}{\begin{eqnarray}}
\newcommand{\eq}{\end{eqnarray}}
\newcommand{\bes}{\begin{subequations}}
\newcommand{\ees}{\end{subequations}}

\def\ben{\begin{eqnarray}}
\def\een{\end{eqnarray}}
\def\ba{\begin{array}}
\def\ea{\end{array}}

\begin{document}
\newcommand{\half}{{\textstyle\frac{1}{2}}}
\allowdisplaybreaks[3]
\def\a{\alpha}
\def\b{\beta}
\def\g{\gamma}\def\G{\Gamma}
\def\d{\delta}\def\D{\Delta}
\def\ep{\epsilon}
\def\et{\eta}
\def\z{\zeta}
\def\t{\theta}\def\T{\Theta}
\def\l{\lambda}\def\L{\Lambda}
\def\m{\mu}
\def\f{\phi}\def\F{\Phi}
\def\n{\nu}
\def\p{\psi}\def\P{\Psi}
\def\r{\rho}
\def\s{\sigma}\def\S{\Sigma}
\def\ta{\tau}
\def\x{\chi}
\def\o{\omega}\def\O{\Omega}
\def\k{\kappa}
\def\pa {\partial}
\def\ov{\over}
\def\br{\\}
\def\ud{\underline}

\newcommand\lsim{\mathrel{\rlap{\lower4pt\hbox{\hskip1pt$\sim$}}
    \raise1pt\hbox{$<$}}}
\newcommand\gsim{\mathrel{\rlap{\lower4pt\hbox{\hskip1pt$\sim$}}
    \raise1pt\hbox{$>$}}}
\newcommand\esim{\mathrel{\rlap{\raise2pt\hbox{\hskip0pt$\sim$}}
    \lower1pt\hbox{$-$}}}
\newcommand{\dpar}[2]{\frac{\partial #1}{\partial #2}}
\newcommand{\sdp}[2]{\frac{\partial ^2 #1}{\partial #2 ^2}}
\newcommand{\dtot}[2]{\frac{d #1}{d #2}}
\newcommand{\sdt}[2]{\frac{d ^2 #1}{d #2 ^2}}    
\newcommand{\aap}{A {\&}A}
\newcommand{\mnras}{MNRAS}

\title{Nonlinear Chaplygin Gas Cosmologies}

\author{P. P. Avelino}
\email[Electronic address: ]{pedro.avelino@astro.up.pt}
\affiliation{Centro de Astrof\'{\i}sica da Universidade do Porto, Rua das Estrelas, 4150-762 Porto, Portugal}
\affiliation{Departamento de F\'{\i}sica e Astronomia, Faculdade de Ci\^encias
da Universidade do Porto, Rua do Campo Alegre 687, 4169-007 Porto, Portugal}
\affiliation{Sydney Institute for Astronomy, School of Physics, A28, The University of Sydney, NSW 2006, Australia}

\author{K. Bolejko}
\email[Electronic address: ]{bolejko@physics.usyd.edu.au}
\affiliation{Sydney Institute for Astronomy, School of Physics, A28, The University of Sydney, NSW 2006, Australia}

\author{G. F. Lewis}
\email[Electronic address: ]{gfl@physics.usyd.edu.au}
\affiliation{Sydney Institute for Astronomy, School of Physics, A28, The University of Sydney, NSW 2006, Australia}

\begin{abstract}

We study the nonlinear regime of Unified Dark Energy models, using Generalized Chaplygin Gas cosmologies as a representative example, and introduce a new parameter characterizing the level of small scale clustering in these scenarios. We show that viable Generalized Chaplygin Gas cosmologies, consistent with the most recent observational constraints, may be constructed for any value of the Generalized Chaplygin Gas parameter by considering  models with a sufficiently high level of nonlinear clustering.

\end{abstract} 
\maketitle

\section{Introduction}

Cosmological observations provide overwhelming evidence for Dark Energy (DE)
\cite{Suzuki:2011hu,Anderson:2012sa,Parkinson:2012vd,Ade:2013zuv}. 
This of course is a subject to several assumptions such as 
that Einstein's General Relativity provides an accurate description of gravity on cosmological scales, and in addition that the Friedmann--Lema\^itre--Robertson--Walker models adequately describe our Universe \cite{Clifton:2011jh,Bolejko:2011jc,Buchert:2011sx}.
Although the simplest DE candidate, a cosmological constant $\Lambda$, is perfectly consistent with current cosmological observations, there is presently no satisfactory explanation for the tiny DE density (over $120$ orders of magnitude smaller than the Planck density) which, nevertheless, appears to account for about $70 \%$ of the total energy density of the Universe at the present time. Hence, despite the simplicity of the cosmological constant, dynamical DE models are arguably better motivated from a theoretical point of view (see, for example, \cite{Copeland:2006wr,Li:2011sd} and references therein).

While DE might explain the observed dynamics of the Universe on cosmological scales, a nonrelativistic dark matter (DM) component is required in the standard cosmological model to account for the observed clustering of large scale structures. The simplest model which incorporates the DM and DE components is the so-called $\Lambda$CDM model, in which the DE and DM roles are played by a cosmological constant $\Lambda$ and Cold Dark Matter (CDM) particles with a negligible free streaming length. The energy components of this model can either be taken as a DE fluid with $p_{\rm DE}=-\rho_{\rm DE}=-\rho_\Lambda$ and a DM fluid with $p_{\rm DM}=0$ or a single Unified DE (UDE) component with $p_{DE}=-\rho_\Lambda$ and arbitrary $\rho > \rho_\Lambda$ (here $\rho$ and $p$ represent the density and pressure, respectively). Hence, the $\Lambda$CDM scenario can be regarded as the simplest example of a UDE realization where the role of DM and DE are played by the same dark fluid \cite{Avelino:2003cf}. Various other interesting candidates for the unification of DM and DE have been proposed in the literature, including the Chaplygin gas and its variations \cite{Kamenshchik:2001cp,Bilic:2001cg,Bento:2002ps}, tachyon field models \cite{Gibbons:2002md,Padmanabhan:2002cp,Bagla:2002yn,Chimento:2003ta,Calcagni:2006ge,Avelino:2010qn,Avelino:2011ey}, and a large variety of Interacting Dark Energy (IDE) models \cite{Bassett:2002fe,Farrar:2003uw,Gumjudpai:2005ry,Boehmer:2008av,Clemson:2011an,Avelino:2012tc}. 

In this paper we shall focus on UDE models in which the UDE fluid is described by a perfect fluid with an isentropic Equation of State (EoS) $p=p(\rho) = w(\rho) \rho$, where $w(\rho)$ is the EoS parameter of the DE, but most of our results will also apply to other IDE models in the strong coupling regime. Despite the very different parameterizations available for $w(\rho)$, all UDE models are characterized by an EoS, satisfying two important properties: i) if the density is much larger than the average density of the Universe at the present time, then $w \sim 0$ (and $c_s \sim 0$, where $c_s$ is the sound speed); ii) if the density is close to the current average density of the Universe then the EoS parameter of the UDE fluid is close to $-1$. A representative example of a family of isentropic UDE models is the Generalized Chaplygin Gas (GCG), characterized by the EoS $p=-A/\rho^{\alpha}$ where $A > 0$ and $0 \le \alpha \le 1$ are constants.

Isentropic UDE models have been claimed to be essentially ruled out due to the late time oscillations of the matter power spectrum inconsistent with observations, except for a small region of parameter space close to the standard $\Lambda$CDM model \cite{Sandvik:2002jz} (for $\alpha < 0$ linear theory would instead predict an exponential blowup of the matter power spectrum). Although the inclusion of baryons in the analysis does lead to less stringent bounds on the GCG parameter $\alpha$ \cite{Beca:2003an}, linear isentropic UDE models have been shown to be tightly constrained by cosmological observations \cite{Alcaniz:2002yt,Dev:2002qa,Makler:2002jv,Bento:2003we,Amendola:2003bz,Cunha:2003vg,Dev:2003cx,Bertolami:2004ic,Biesiada:2004td,Wu:2006pe,Wu:2007bv,Gorini:2007ta,Lu:2008zzb,Fabris:2010yh,Xu:2010zzb,Lu:2010zzj,Fabris:2011wk,Xu:2012ca,Wang:2013qy} (see also \cite{Reis:2003mw,Zimdahl:2005ir,Bilic:2006cp,Bertacca:2010ct} for a discussion of nonisentropic UDE models).

The effect of small scale nonlinearities has been recognized as having a strong potential impact on the large scale evolution of the Universe, in particular in the context of UDE scenarios,  \cite{Avelino:2003ig,Bilic:2003cv,Beca:2005gc,Avelino:2007tu,Avelino:2008cu,Roy:2009cj,DelPopolo:2013bpa}. However, it has been argued that, in the case of the Chaplygin gas, nonlinear effects would be too small to significantly affect the above linear results \cite{Bilic:2003cv} (see also \cite{Avelino:2003ig}). This conclusion relied on the assumption of a constant spectral index of scalar gaussian fluctuations as well as an EoS for the Chaplygin gas whose form remains unchanged both at large densities and small scales. These are very strong assumptions, given the relatively weak constraints on the scalar spectral index on nonlinear scales (wavenumbers $k \gsim 3 \, {\rm Mpc}^{-1}$) and the expectation that the isentropic perfect fluid approximation might break at sufficiently large densities or small scales.

In this paper we relax these assumptions and model the effect of the small scale nonlinearities on the background evolution of the Universe using a single parameter $\epsilon$, representing the fraction of the UDE density which has become nonlinear due to the gravitational collapse into UDE halos. We show that, for $\epsilon$ close to unity, the linear theory results no longer hold and that the backreaction of the small scale structures on the large scale evolution of the Universe may render the Chaplygin gas model virtually indistinguishable from the $\Lambda$CDM scenario for all possible values of the GCG parameter $\alpha$.

In this paper we shall use units where $8 \pi G/3=1$.

\section{Homogeneous Unified Dark Energy Models\label{homog}}

In this section we shall consider a perfectly homogeneous and isotropic universe made up of mainly two components: a baryonic component of density $\rho_b$ and negligible pressure and a UDE fluid component of density $\rho$ and pressure $p$. Energy-momentum conservation implies
\be
(\ln \rho)'+3(1+w)=0\,, \label{enec}
\ee
where $w=p/\rho$ is the EoS parameter of the UDE fluid and a prime represents a derivative with respect to $\ln a$. By setting $(\ln \rho)'=0$ one obtains $w=-1$, usually signaling a minimum density in a cosmological context. The complete solution to this equation can be written in the general form
\be
\rho=\rho_0 \exp \left(3 \int_{\ln a}^{\ln a_0} (1+w(x)) dx\right)\,,
\label{rhoev}
\ee
where the subscript $0$ refers to the present time time $t_0$. For simplicity we shall take $a_0=1$, so that $\ln a_0=0$.

We shall consider the GCG, characterized by the EoS parameter
\be
w \equiv \frac{p}{\rho}=-\frac{A}{\rho^{1+\alpha}}\,,
\ee
where $A$ is a positive constant and $0 \le \alpha \le 1$, as a simple representative example of a family of UDE fluids; for $\alpha=0$ the GCG model is completely equivalent to $\Lambda$CDM. However, the best motivated model from the GCG family is characterised by $\alpha=1$, due to an interesting connection to string theory \cite{Ogawa:2000gj}.

The evolution of the (homogeneous) GCG density $\rho$ as a function of the scale factor $a$ is given by
\be
\rho=\rho_0 \left[(1-{\bar A})a^{-3(1+\alpha)}+{\bar A}\right]^{\frac{1}{1+\alpha}}\,,
\ee
where $\bar A=A/\rho_0^{1+\alpha}$ and
\be
\rho_{\rm min}=\rho_0 {\bar A}^{\frac{1}{1+\alpha}}\,,
\ee
is the minimum density, for which $w=-1$. 

In a flat Universe the Hubble parameter $H \equiv {\dot a}/a$ and the deceleration parameter are respectively given by
\be
H^2=\rho + \rho_b\,,
\ee
and
\be
q \equiv - (\ln H)' -1= \frac{1}{2} \left(\Omega\left(1+3w\right)+\Omega_b\right)\,,
\ee
where a dot represents a derivative with respect to the physical time, $\Omega=\rho/\rho_c$, $\Omega_b=\rho_b/\rho_c$ and $\rho_c$ is the critical density (note that, in a flat universe, $\Omega+\Omega_b=1$).

The transition from a decelerating to an accelerating regime occurs when $q=0$. For $\Omega_b=0$ the transition would occur at the scale factor
\be
a_{q=0}=\left(\frac{1-{\bar A}}{2{\bar A}}\right)^{\frac{1}{3(1+\alpha)}}\,.
\ee

\section{Backreaction effects \label{inhomog}}

In this section we shall study the backreaction of small scale nonlinearities on the large scale evolution of the Universe in UDE scenarios, using the GCG as representative family of UDE models. In order to make the problem more tractable, we shall assume that the distribution of the GCG component in a large comoving volume $V$ of the Universe is essentially composed of two types of regions: i) collapsed regions `+' with a GCG characteristic density much greater than the average GCG density $\rho$ and consequently with a very small pressure and volume (which, for simplicity, we shall assume to be zero in the estimation of the EoS parameter of the GCG); ii) underdense regions `-' with densities smaller than $\rho$ which occupy most of the volume of the Universe.

The average fraction of the total GCG energy $E$ which is incorporated into collapsed objects (with total energy $E_+$) in a comoving region of the Universe of comoving volume $V$ will be parametrized by
\be
\epsilon=E_+/E\,,
\ee
which quantifies the level of small scale clustering.
The contribution of the collapsed regions to the average density of the Universe is then given by
\be
\rho_+=E_+/V=\epsilon \rho\,.
\ee
On the other hand, the average density in underdense regions is just
\be
\rho_-=\frac{E_-}{V}=\frac{E-E_+}{V}=(1-\epsilon)\rho\,.
\ee

Eq. (\ref{enec}) remains valid in the presence of small scale nonlinearities in the GCG component. However, in this case, the contribution to the pressure comes solely from the underdense regions so that
\be
w=\frac{p_-}{\rho}=\frac{\rho_-}{\rho}\frac{p_-}{\rho_-}=(1-\epsilon)w_-\,,
\ee
where $w_-=p_-/\rho_-$ can be identified as the effective DE EoS parameter.

\begin{figure}[t]
	\centering
	\includegraphics[width=9.0cm]{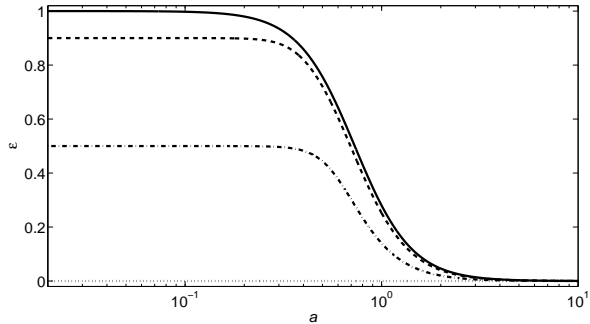}
\caption{Evolution of the parameter $\epsilon$ with the scale factor $a$ for four different models: $\epsilon_i=0$, $\alpha=1$ (dotted line), $\epsilon_i=0.5$, $\alpha=1$ (dot-dashed line), $\epsilon_i=0.9$, $\alpha=1$ (dashed line) and $\epsilon_i=[1+{\bar A}a_i^3/(1-{\bar A})]^{-1}$, $\alpha=0$ (solid line). Note that the solid line corresponds to the $\Lambda$CDM case. \label{fig-1}}
\end{figure}

Having added a new parameter $\epsilon$ characterizing the level of small scale clustering in UDE models, one also needs to specify its evolution. Here we shall consider a simple model where $E_+$ remains fixed. Given that $E \propto \rho a^3$ one has
\be
\epsilon=\frac{\epsilon_0 \rho_0}{\rho a^3}\,.
\ee
At early times $\rho \propto a^{-3}$ and $\epsilon$ becomes a constant ($\epsilon \to \epsilon_i$). The generalization to an arbitrary evolution of $E_+$ is straightforward (in more realistic models $E_+$ is expected to increase with $a$). 

In this paper we shall assume that $\Omega_{b0}=0.0487$ and $H_0=67.3 \, {\rm km \, s^ {-1} \, Mpc^ {-1}}$, in agreement with the latest Planck results \cite{Ade:2013zuv}. We shall also use the estimate of the matter density parameter obtained by the Planck collaboration, $\Omega_{m0}=0.315$, to fix the baryonic and GCG fractional densities at early times (into the matter dominated era) to be, respectively, equal to
\be
\Omega_{bi}=\frac{\Omega_{b0}}{\Omega_{m0}}=0.155\,, \qquad \Omega_{bi}=1-\Omega_i\,.
\ee
By choosing these values for the cosmological parameters, we ensure that at recombination our GCG model  is fully consistent with the Planck Cosmic Microwave Background (CMB) constraints (note that at early times the GCG behaves as CDM).

Fig. 1 shows the evolution of the parameter $\epsilon$ for four different models: $\epsilon_i=0$, $\alpha=1$ (dotted line), $\epsilon_i=0.5$, $\alpha=1$ (dot-dashed line), $\epsilon_i=0.9$, $\alpha=1$ (dashed line) and $\epsilon_i=[1+{\bar A}a_i^3/(1-{\bar A})]^{-1}$, $\alpha=0$ (solid line). Although in the later case, with $\alpha=0$, the actual value of $\epsilon_i$ is not relevant for the evolution of the average GCG  density $\rho$, since it corresponds to the $\Lambda$CDM limit of the GCG scenario, it has been chosen to account for a simple decomposition of the UDE fluid into matter and a cosmological constant. As previously mentioned, the present time is achieved when  $\Omega_b=0.0487$, which may happen at different values of $\epsilon_i$ in different models (as expected, these differences completely vanish in the $\epsilon \to 1$ limit). 

\begin{figure}
	\centering
	\includegraphics[width=9.0cm]{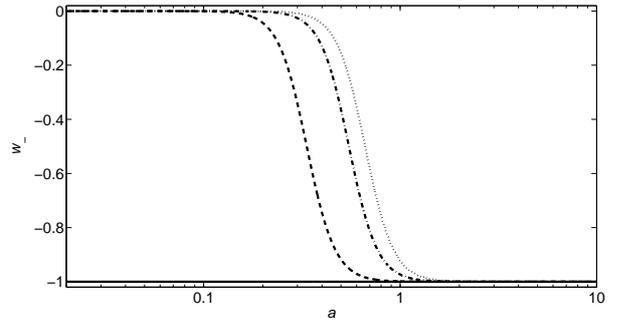}
\caption{Evolution of the effective DE EoS parameter $w_-$ with the scale factor $a$ for the models considered in Fig. 1. \label{fig-2}}
\end{figure}

The evolution of $\epsilon=E_+/E$ with the scale factor $a$ in Fig. 1 shows that $\epsilon$ tends to a constant value $\epsilon_i$ for $a \ll 1$, evolving rapidly towards zero for $a \gg 1$. This is the expected behavior since $E$ is roughly a constant during the matter dominated era and grows proportionally to $a^3$ in the DE dominated era. Thus, for fixed $E_+$, the asymptotic evolution of $\epsilon$ with the scale factor is just $\epsilon={\rm constant}$ into the matter dominated era and $\epsilon \propto a^{-3}$ into the DE dominated era.

Fig. 2 shows the evolution of the effective DE EoS parameter $w_-$ with $a$ for the models considered in Fig. 1. Except for the $\alpha=0$ case (the $\Lambda$CDM limit of the GCG), the value of $w_-$ smoothly interpolates from $w_-=0$ into the matter dominated era to $w_-=-1$ into the DE era, with the scale factor at the transition being controlled by the parameter $\epsilon_i$. As expected, Fig. 2 shows that as the amount of small scale clustering is increased (by increasing $\epsilon_i$) the transition from a cosmological constant to a CDM behaviour occurs at larger and larger redshifts.

Although the evolution of the DE EoS parameter $w_-$ cannot in general be found analytically, a simple fit to a numerical solution can nevertheless be found in the constant $E_+$ case studied in the present paper. The transition from $w_-=0$ to $w_-=-1$ occurs for $\rho_- \sim \rho_{\rm min}$. Taking into account that during the $w_-=0$ phase 
\be
\rho_-=(1-\epsilon)\rho \propto (1-\epsilon_i)a^{-3}\,,
\ee
one finds that the scale factor $a_{\rm tr}$ at the transition between the two phases is roughly proportional to $(1-\epsilon_i)^{1/3}$. Taking into account that, if $\epsilon_i = 0$, the transition from $w_-=0$ to $w_-=-1$ occurs for a scale factor approximately equal to $[(1-{\bar A})/{\bar A}]^{1/(3(1+\alpha))}$ one finds that
\be
a_{\rm tr} = (1-\epsilon_i)^{1/3} [(1-{\bar A})/{\bar A}]^{\frac{1}{3(1+\alpha)}}\,.
\ee
Note that $a_{\rm tr} \to 0$ for $\epsilon_i \to 1$, as expected in the $\Lambda$CDM limit of the Chaplygin gas. A fit which takes the above scaling into account is given by
\be
w_-=-\frac{{\bar A}}{Ba^{-3(1+\alpha)}+{\bar A}}\,,
\ee
with $B=(1-{\bar A})(1-\epsilon_i)^{1+\alpha}$. For $\epsilon_i \sim 0$ one has $B \sim 1-{\bar A}$ and the background result is recovered. On the other hand, $B=0$ in the $\epsilon \to 1$ limit so that $w_-=-1$ for all values of the scale factor $a$, as expected of a cosmological constant.

\begin{figure}
	\centering
	\includegraphics[width=9.0cm]{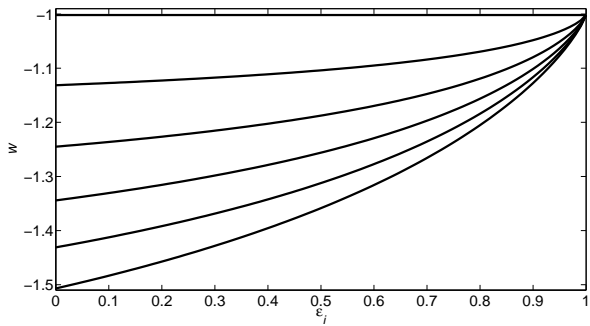}
\caption{The lines show the correspondence between the values of $w$ and $\epsilon_i$, for standard quintessence and GCG models with the same angular diameter distance to the last scattering surface, and sound horizon as constrained by Planck. The assumed values of $\alpha$ of the GCG model are, from top to bottom, $\alpha=0,0.2,0.4,0.6,0.8,1$, respetively. \label{fig-3}}
\end{figure}


\section{Cosmological observations\label{angdis}}

In order to have agreement with the Planck results \cite{Ade:2013zuv}, one needs to ensure that the angular diameter distance to the last scattering surface 
\be
d_\theta(z_{\rm rec})=\frac{1}{1+z_{\rm rec}}\int_0^{z_{\rm rec}} \frac{dz}{H(z)}\,,
\ee
and that the sound horizon is compatible with the latest CMB observations. Here $z=1/a-1$ is the redshift and the subscript `rec' represents recombination. In the following, we shall find the standard quintessence model with constant $w$ (and $\Omega_{b0}=0.0487$, $\Omega_{m0}=0.315$ and $H_0=67.3 \, {\rm km \, s^ {-1} \, Mpc^ {-1}}$) which has the same distance to the last scattering surface as that obtained for the GCG model, parametrised by $\alpha$ and $\epsilon_i$, with the corresponding choices for the cosmological parameters.

The lines in Fig. 3 show the correspondence between the values of $w$ and $\epsilon_i$, for standard quintessence and GCG models with the same angular diameter distance to the last scattering surface. The values of $\alpha$ corresponding to the different lines are $\alpha=0,0.2,0.4,0.6,0.8,1$ (from top to bottom, respectively). It is clear that the $\Lambda$CDM limit is recovered both for $\alpha=0$ (irrespectively of the value of $\epsilon_i$) and for $\epsilon_i \to 1$ (irrespectively of the value of $\alpha$). This result implies that, independently of the value of $\alpha$, GCG models can be made compatible with current observational constraints as long as the level of nonlinear clustering is high enough. For $\epsilon > 0.9$, the corresponding value of $w$ is well within the latest observational uncertainties \cite{Ade:2013zuv}  for all $0 \le \alpha \le 1$.

\begin{figure}
\begin{center}
\includegraphics[width=9.0cm]{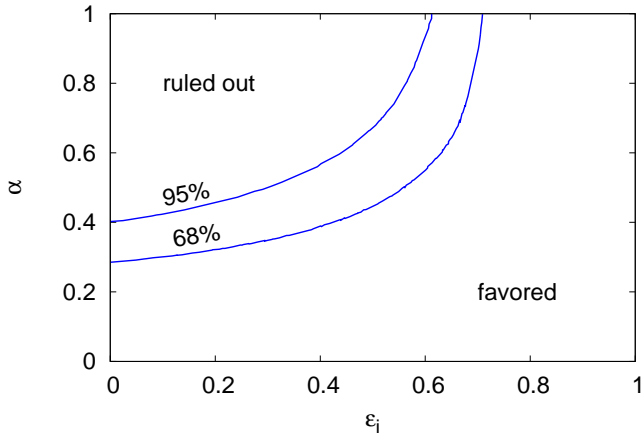}
\caption{68\% and 95\% confidence regions based on the supernova observations.
The colours correspond to the likelihood (${\cal L}/{\cal L}_{max}$).}
\label{fsn}
\end{center}
\end{figure}

The values of $\alpha$ and $\epsilon$ may also be constrained using supernova data. The likelihood 
 \[ {\cal L} = {\rm e}^{-\chi^2/2}, \]
is given by
\begin{equation}
 \chi^2 = (\vec{M} - \vec{D})^T C^{-1} (\vec{M} - \vec{D}),
\end{equation}
where  $\vec{M}$ is the distance modulus
\begin{equation}
 M_i = 5 \log_{10} \left[ (1+z_i) \int\limits_0^{z_i} \frac{ {\rm d} \tilde{z} } { H(\tilde{z}) } \right]  +25,
\end{equation}
and $\vec{D}$ and $C$ are respectively the distance modulus and covariance matrix of
the Union2.1 data set \cite{Suzuki:2011hu}.
The 68\% and 95\% confidence regions are presented in Fig.~\ref{fsn}.

As seen, now with nonlinear clustering effect included, i.e. when $\epsilon \to 1$,
the Chaplygin Gas ($\alpha =1$) is consistent with supernova constraints.
Without taking into account the nonlinear clustering effects 
($\epsilon = 0$) the Chaplygin Gas model would be ruled out.
In the next section we show that these nonlinear clustering effects also improve
the status of the Chaplygin model in regards to the growth of cosmic structures.

\section{Evolution of density perturbations\label{denper}}

Although a detailed study of the evolution of density perturbations is outside the scope of this paper, we shall atempt to describe its main features. At late times a large sound speed prevents the growth of density perturbations of the `-' component of the GCG. This happens, at different times on different scales, when the comoving sound horizon ($\sim c_{s-}/(Ha)$) multiplied by the comoving wavenumber ($k$) becomes of order unity. At early times, into the matter dominated era, $H \sim H_0 a^{-3/2}$ and $c_{s-}^2=- \alpha w_- \sim \alpha a^{3(\alpha+1)}/(1-\epsilon_i)^{1+\alpha}$. Hence, a fluctuation of wavenumber $k$ will stop growing roughly at the scale factor
\be
a_k \sim \alpha^{-\frac{1}{4+3 \alpha}} (1-\epsilon_i)^{\frac{1+\alpha}{4+3 \alpha}} \left(\frac{k}{H_0}\right)^ {-\frac{2}{4+3 \alpha}}\,.
\ee

On the other hand, into the matter dominated era, the linear evolution of density perturbations of the `+'  GCG component is described by the equation (see, for example, \cite{Beca:2003an})
\be
\delta_+'' + (2+\zeta)\delta_+' - \frac32\left[\Omega_+ \delta_+ +  (1+3 c_{s-}^ 2)\Omega_- \delta_- + \Omega_b \delta_b \right]=0\,, \label{denev}
\ee
with $\zeta=H'/H=-3/2$. For the sake of simplicity, we shall assume that $\delta_+=\delta_b$ and absorb the baryons in the `+' component of the GCG, so that $\Omega_+ \sim \epsilon_i$ at early times.

If $c_s \ll 1$ and $a \ll a_k$  one obtains the standard result for the linear growing mode DM perturbations in the matter dominated era, which is $\delta_+ \propto \delta_- \propto a$. On the other hand, for $a \gg a_k$ the fluctuations in the `-' component of the GCG will be negligible ($\delta_- \sim 0$) and Eq. (\ref{denev}) becomes
\be
\delta_+''+\frac12 \delta_+' -\frac32 \epsilon_i \delta_+=0\,.
\ee
In this case, the growing mode solution is given by
\be
\delta_+ \propto a^{1-\frac{3(1-\epsilon_i)}{5}}\,,
\ee
for $1-\epsilon_i \ll 1$.
This results in a smaller growth factor with respect to the standard case by
\be
f=a_k^{-\frac{3(1-\epsilon_i)}{5}} \sim 1+\frac35 \ln a_k (1-\epsilon_i)\,, 
\ee
where the approximation is valid if $|f-1| \ll1$.  For $\epsilon_i > 0.9$ this condition is satisfied for $k \lsim 0.3 \, {\rm Mpc}^{-1}$ (and even on smaller nonlinear scales). This implies that the late time oscillations of the matter power spectrum which plagues linear GCG models can be avoided if the level of nonlinear clustering is sufficiently high, thus rendering the model consistent with observations.

\section{Conclusions \label{conc}}

In this paper we have parametrised the effect of UDE nonlinear clustering on the dynamics of the Universe. We have shown that cosmological scenarios in which the DM and DE roles are played by a single UDE fluid may be reconciled with the latest observational results, provided there is a high level of nonlinear clustering  of the UDE component. Although we have focused on the GCG as a concrete example, our main results are  expected to hold in general for UDE models.

\begin{acknowledgments}
We thank Vasco Ferreira for useful comments. The work of P.P.A. was supported by Fundação para a Ciência e a Tecnologia (FCT) through the Investigador FCT contract No. IF/00863/2012 and POPH/FSE (EC) by FEDER funding through the program Programa Operacional de Factores de Competitividade - COMPETE, and by Grant No. PTDC/FIS/111725/2009 (FCT). G.F.L. acknowledges support through an Australian Research Council Discovery Project (DP130100117).
\end{acknowledgments}

\bibliography{UDE}

\end{document}